\documentclass[12pt]{article}
\usepackage{amsfonts}
\usepackage{amsmath}
\usepackage{amssymb}
\usepackage{mathptmx}
\usepackage{epsfig}
\usepackage{amsmath}
\usepackage{amssymb}

\def\half{\textstyle\frac12}
\def\3half{\textstyle\frac32}
\begin{document}
\title{Revisiting $B\to \phi\pi$ Decays in the Standard Model  }
\author{Ying Li\footnote{e-mail: liying@ytu.edu.cn.}
 \\
{\small \it  Department of Physics, Yantai University, Yantai
264005, Peoples' Republic of China} \\
Cai-Dian L\"u, Wei Wang\\
{\small \it  Institute of High Energy Physics, CAS, Beijing 100049,
Peoples' Republic of China} } \maketitle

\begin{abstract}
In the standard model (SM), we re-investigate the rare decay $B\to
\phi\pi$, which has been viewed as an ideal probe to detect the new
physics signals beyond the SM. Contributions in the naive
factorization method, the radiative corrections, the long-distance
contributions, and the contributions due to the $\omega$-$\phi$
mixing are taken into account. We find that the tiny branching ratio
in the naive factorization can be dramatically enhanced by the
radiative corrections and the $\omega$-$\phi$ mixing effect, while
the long-distance contributions are negligibly small. Assuming the
Cabibbo-Kobayashi-Maskawa angle $\gamma=(58.6\pm 10)^\circ$ and the
mixing angle $\theta= -(3.0\pm 1.0)^\circ$, we obtain the branching
ratios of $B\to \phi\pi$ as $ {\rm Br}(B^\pm\to \phi \pi^\pm)= (3.2
^{+0.8-1.2}_{-0.7+1.8}) \times 10^{-8}$ and $ {\rm Br}(B^0 \to \phi
\pi^0) = (6.8 ^{+0.3-0.7}_{-0.3+1.0}) \times 10^{-9}$. If the future
experiment reports a branching ratio of order $10^{-7}$ for $B^-\to
\phi \pi^-$ decay, it may not be a clear signal for any new physics
scenario. In order to discriminate the large new physics
contributions and those due to the $\omega$-$\phi$ mixing, we
propose to measure the ratio of branching fractions of the charged
and neutral B decay channel.  We also study the direct $CP$
asymmetries of these two channels, and the results are about
$(-8.0^{+0.9+1.5}_{-1.0-0.1})\%$ and
$(-6.3^{-0.5+2.5}_{+0.7-2.5})\%$ for $B^\pm\to \phi \pi^\pm$ and
$B^0 \to \phi \pi^0$, respectively.
\end{abstract}

\begin{figure}[tb]
\begin{center}
\psfig{file=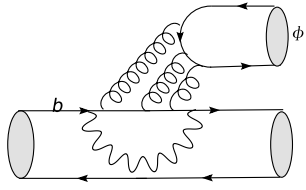,width=4.0cm,angle=0}
\end{center}
\caption{Hairpin diagrams for $B\to \phi\pi$ decays}\label{Feyn:O1}
\end{figure}

\newpage
$B$ meson decays provide valuable information on the flavor
structure of the weak interactions so that they can be used to
precisely test the standard model (SM) and to search for the
possible signals of the new physics beyond the SM. Charmless
two-body nonleptonic decay processes, such as $B\to \phi\pi$, are of
great interests, since the branching fractions are very small. The
experimentalists have reported the following
measurements~\cite{Aubert:2006nn}:
\begin{eqnarray}
 { BR}(B^-\to \phi\pi^-)&=&(-0.04\pm0.17)\times 10^{-6},\nonumber\\
 { BR}(\bar B^0\to \phi\pi^0)&=&(0.12\pm0.13)\times 10^{-6}
\end{eqnarray}
at $68\%$ probability, while the upper bounds at $90\%$ probability
are given as:
\begin{eqnarray}
 { BR}(B^-\to \phi\pi^-)&<&2.4\times 10^{-7},\label{eq:upperbound1}\\
 { BR}(\bar B^0\to \phi\pi^0)&<&2.8\times
 10^{-7}.\label{eq:upperbound2}
\end{eqnarray}
On the theoretical side, since these decay modes are absent from any
annihilation diagram contribution,  calculations of  hadronic matrix
elements are quite reliable, and these decays have been analyzed in
the SM by different groups \cite{phism,Ali:1998eb}. In the SM, these
channels are highly suppressed for several reasons. At the quark
level, these decays proceed via $b\to d \bar ss$, which is a flavor
changing neutral current (FCNC) process. The FCNC transition is
induced by the loop effects and the relevant Wilson coefficients are
very small. Secondly, the Cabibbo-Kobayashi-Maskawa (CKM) matrix
element for this transition $V_{tb}V_{td}^*$ is tiny. Finally, in
order to produce a $\phi$-meson from the vacuum, at least three
gluons are required which suppresses these  channels further.
Feynmann diagrams for these decays are often referred to the hairpin
diagram (the last reference in Ref.\cite{phism}), which are shown in
Fig.~\ref{Feyn:O1}. Because of the tiny branching ratio in the SM,
$B\to \phi\pi$ is usually considered as an ideal place to search for
the possible new physics scenarios~\cite{phinewphysics}.

However, before we turn to the new physics scenario, it is logical
to investigate all possible contributions in the SM: contributions
in the naive factorization, radiative corrections (vertex
corrections and the hard spectator diagram),  long-distance
contributions such as rescattering from $B\to KK^*$ decays and
contributions due to the $\omega$-$\phi$ mixing.  The motif of this
letter is to investigate the possibility of the enhancement of $B\to
\phi\pi$ decays in the SM.

The $\Delta B = 1$ effective weak Hamiltonian in SM is given by
\cite{Buchalla:1996vs}:
\begin{equation}\label{Heff}
   { H}_{\rm eff} = \frac{G_F}{\sqrt2} \sum_{p=u,c} \!
   \lambda_p \bigg( C_1\,Q_1^p + C_2\,Q_2^p
   + \!\sum_{i=3}^{ 10, 7\gamma, 8g}\! C_i\,Q_i   \bigg) + \mbox{h.c.} \,,
\end{equation}
where $\lambda_p=V_{pb}V_{pd}^{*}$. $Q_{1,2}^p$ are the left-handed
current--current operators arising from $W$-boson exchange,
$Q_{3,\dots, 6}$ and $Q_{7,\dots, 10}$ are QCD and electroweak
penguin operators, and $Q_{7\gamma}$ and $Q_{8g}$ are the
electromagnetic and chromomagnetic dipole operators, respectively.
Their explicit expressions can be found in, e.g.,
Ref.~\cite{Buchalla:1996vs}.

The physics above the scale $m_b$ in the $B$ meson weak decays have
been incorporated into the Wilson coefficients of the effective
Hamiltonian and the task left is to evaluate the matrix element of
each four-quark operator. The simplest way is to decompose it into
two simpler parts: one is the decay constant of the emitted meson;
the other part is the $B$-to-light form factor which only contains
one light meson in the final state. Both of these two parts can be
directly extracted from the experimental data, or evaluated from
some nonperturbative method such as the Lattice QCD and QCD Sum
Rules. In this approach, often referred as the naive
factorization\cite{Bauer:1984zv}, the decay amplitudes can be
written as:
\begin{eqnarray}
& &A_{B^-\to \pi^-\phi}^{NF} =\sqrt{2}\,A_{\bar B^0\to
\pi^0\phi}^{NF} =A_{\pi\phi}\sum_{p=u,c} \lambda_p (a_3+a_5 -
\frac{1}{2}
 a_7-\frac{1}{2}a_9),
\end{eqnarray}
where
\begin{eqnarray}
& & A_{\pi\phi}=-i \sqrt 2G_Fm_\phi f_\phi (\epsilon^*_\phi \cdot
p_B) F_+^{B \pi},
\end{eqnarray}
and $a_i$ is the Wilson coefficient combination as  defined in
Ref.\cite{Ali:1998eb}. In the naive factorization, predictions for
the branching ratios are given as:
\begin{eqnarray}\label{naivedirectratio}
{\rm Br}(B^\pm \to \phi \pi^\pm) &=& 9.0 \times 10^{-10}, \nonumber \\
{\rm Br}(B^0 \to \phi \pi^0) &=& 4.1 \times 10^{-10}.
\end{eqnarray}
In the calculation, we have used $f_\phi=0.22$ GeV. The particle
masses and lifetime of the $B$ meson are taken from
\cite{Amsler:2008zz}. The value of the form factor at zero recoil is
taken as $F_+^{B \pi}$=0.25. The value of the CKM matrix elements
used are taken from \cite{Amsler:2008zz}:
\begin{eqnarray}
 |V_{ub}|=0.0039\;,\;|V_{ud}|=0.974,\;,\; |V_{cb}|=0.0422\;,\;|V_{cd}|=0.226,
\end{eqnarray}
and the phase $\gamma$~associated with $V_{ub}$ as $58.6^{\circ}$.
Compared with Eq.~\eqref{eq:upperbound1} and \eqref{eq:upperbound2},
we can see the results in the naive factorization are far below the
experimental upper bound. The tiny branching ratios are due to the
cancelation of the Wilson coefficients $C_3,C_4,C_5,C_6$. This
cancelation also reflects the fact that $\phi$ can only be produced
by at least three gluons. The RG evolved Wilson coefficients at the
scale $m_b$ have contains the multi-loop contributions above the
scale $m_b$ which give small branching fractions to $B\to\phi\pi$
decays. Below this scale, the radiative corrections may provide
sizable contributions. In the QCD factorization (QCDF)
approach~\cite{Beneke:1999br,Beneke:2003zv}, the hadronic matrix
elements of local operators $Q_i$ can be written as
\begin{eqnarray}\label{fact}
\langle\pi(p)\phi(q)|Q_i|\bar B(p)\rangle&=& F_+^{B\to\pi}\int_0^1
dv\, T^I
(v)\Phi_{\phi}(v) \nonumber \\
&&+\int_0^1 d\xi dudv\,
T^{II}(\xi,u,v)\Phi_B(\xi)\Phi_{\pi}(u)\Phi_{\phi}(v)
\end{eqnarray}
where $\phi_M$ ($M=\phi, \pi, B$) are light-cone distribution
amplitudes of the meson $M$, $T^I_{i}$ and $T^{II}_i$ are hard
scattering  kernels. To be more specific, the Wilson coefficient
combination $a_3+a_5$ is replaced by the $\alpha_3^p$, while
$a_7+a_9$ is replaced by the $\alpha_{3EW}^p$ which has been defined
in Ref.\cite{Beneke:2003zv}. For the numerical evaluation, we use
the input parameters as given in the QCD factorization approach
\cite{Beneke:2003zv}. With these input parameters,  branching ratios
are obtained as:
\begin{eqnarray}\label{directratio}
{\rm Br}(B^\pm \to \phi \pi^\pm) &=& 1.1 \times 10^{-8}, \nonumber \\
{\rm Br}(B^0 \to \phi \pi^0) &=& 5.2 \times 10^{-9} .
\end{eqnarray}
Compared with  results in the naive factorization approach, we find
that the branching ratios are enhanced by a factor of $12.3$. We
also find that our results are larger than those evaluated in the
QCDF approach in Ref.\cite{phinewphysics,Beneke:2003zv}. The reason
is that we have chosen the factorization scale for the
hard-spectator diagram as $\mu=1$ GeV. If we chose the factorization
scale is $\mu=2.1$ GeV, the branching ratio will be reduced by a
factor of 2. The difference caused by the factorization scale
characterizes the size of the subleading corrections for the hard
scattering diagrams.

Despite  of the perturbative contributions,  $B\to \phi\pi$ decays
also receive some nonperturbative corrections: $B\to K^{(*)}
K^{(*)}$ then $K^{(*)} K^{(*)} \to \phi \pi$ through exchanging a
$K^{(*)}$ meson, which is also called final state interaction. In
the $m_b\to \infty$ limit, the FSI is power suppressed and believed
to vanish. Since the $b$ quark mass is finite, the FSI is not zero
and the $t$-channel FSI has been modeled as the
one-particle-exchange picture~\cite{Cheng:2004ru}. As an example, we
will study the FSI effect from the $B^-\to K^{*-}K^0$ decays.  The
short distance contribution to the $B^-\to K^{*-}K^0$ is given as:
\begin{eqnarray}
 A(B^-\to K^{*-}K^0)&=&  -i\frac{G_F}{\sqrt 2}f_{K}A_0^{BK^*} (2m_{K^*}\epsilon^*_{K^*}\cdot
 p_B)\sum_p\lambda_p[
  \alpha_4^p-\frac{1}{2}\alpha_{4,EW}^{p}]
\end{eqnarray}
The long-distance contribution to $B^-\to \phi\pi^-$ is given as:
\begin{eqnarray}
  A_{abs}&=&-i\frac{G_F}{\sqrt 2}f_{K}A_0^{BK^*}\sum_p\lambda_p[
  \alpha_4^p-\frac{1}{2}\alpha_{4,EW}^{p}]
   \int_{-1}^{1}
 \frac{|\vec p_1|d\cos\theta}{16\pi m_B} 4m_{K^*}g_{K^*K\pi} g_{\phi KK} \nonumber\\
 &\times&(-p_B\cdot p_3+ \frac{p_B\cdot p_1{p_1\cdot p_3}}{m_{K^*}^2})
 \times  \frac{E_2|\vec p_4|-E_4 |\vec p_2|\cos\theta}{m_\phi} \times
 \frac{F(t,m^2)}{t-m^2},
\end{eqnarray}
where $p_1,p_2,p_3,p_4$ denotes the momentum of the
$K^{*-},K^0,\pi^-,\phi$ mesons, respectively. $\theta$ is the angle
between the momenta $\vec p_1$ and $\vec p_3$. The coupling
constants $g_{\phi KK}$ and $g_{K^* K\pi}$ can be determined through
the experimental data on $\phi\to KK$ and $K^*\to K\pi$
decays~\cite{Amsler:2008zz}, and we get $g_{\phi KK}=4.51$ and
$g_{K^*K\pi}=4.86$. Because of the small branching ratios (of  order
$10^{-7}$) of $B\to KK^*$ decays~\cite{data}, the long-distance
contributions to $B\to \pi\phi$ decays are not expected to give
sizable corrections. The numerical results also show that these
contributions are negligibly small.
\begin{figure}[hbtp]
\begin{center}
\includegraphics[width=0.5\textwidth]{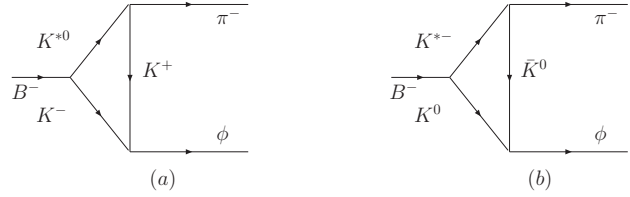}
\caption{Feynman diagrams of the final state interactions }
\label{Feyn:FSI}
\end{center}
\end{figure}

All the above investigations are based on the hypothesis that
$\omega-\phi$ are ideally mixing: $\omega=\frac{u\bar u+d\bar
d}{\sqrt{2}}$ and $\phi=s\bar s$. But generally, the $\omega$ and
$\phi$ can mix with each other via the SU(3) symmetry breaking
effect.  With the aid of a mixing angle $\theta$, one can
parameterize the $\omega-\phi$ mixing, so that the physical $\omega$
and $\phi$ are related to the two states $n\bar n=\frac{u\bar
u+d\bar d}{\sqrt{2}}$ and $ s\bar s$
\begin{eqnarray}\label{mixing}
\left(
  \begin{array}{c}
     \omega\\
    \phi
  \end{array}\right)=\left(
                \begin{array}{cc}
                  \cos\theta & \sin\theta \\
                  -\sin\theta & \cos\theta \\
                \end{array}
              \right)\left(\begin{array}{c}
      n\bar n\\
    s\bar s
  \end{array}
\right)
\end{eqnarray}
Recent studies within the  chiral perturbative theory imply a mixing
angle of $\theta= -(3.4\pm 0.3)^\circ$~\cite{Benayoun:1999fv}, while
the most recent treatment implies an energy-dependent mixing which
varies from $ -0.45 ^\circ$ at the $\omega$ mass to $ -4.64 ^\circ$
at the $\phi$ mass \cite{Benayoun:2007cu}.  Although the $n\bar n$
component in the $\phi$ meson is tiny, it may sizably contribute to
some observables such as branching ratio and direct $CP$ violation
parameters of the rare decays such as $B\to \phi\pi$
\cite{Gronau:2008kk}.

As the $n\bar n$ component concerned, both the emission and
annihilation topologies contribute to these decays. Therefore, not
only penguin operators but also tree operators should be taken into
account. For the $n\bar n$ part, the decay amplitudes are given:
\begin{multline}
\sqrt2\,{  A}_{B^-\to\pi^-\phi}^{n\bar n}
  = A_{\phi\pi} \sum_{p=u,c} \lambda_p\Big[
    \delta_{pu}\,(\alpha_2 + \beta_2 )
    + 2\alpha_3^p + \alpha_4^p + \half\alpha_{3,{\rm EW}}^p
    - \half\alpha_{4,{\rm EW}}^p
    +\, \beta_3^p + \beta_{3,{\rm EW}}^p
     \Big] \\
  + A_{\phi\pi} \sum_{p=u,c} \lambda_p\Big[ \delta_{pu}\,(\alpha_1 + \beta_2)
    + \alpha_4^p + \alpha_{4,{\rm EW}}^p + \beta_3^p
    + \beta_{3,{\rm EW}}^p \Big],\label{xx1}
\end{multline}
\begin{multline}
   -2\,{  A}_{\bar B^0\to\pi^0\phi}^{n\bar n}
   = A_{\phi\pi} \sum_{p=u,c} \lambda_p\Big[
    \delta_{pu}\,(\alpha_2 - \beta_1 )
    + 2\alpha_3^p + \alpha_4^p + \half\alpha_{3,{\rm EW}}^p
    - \half\alpha_{4,{\rm EW}}^p+\, \beta_3^p -\half\beta_{3,{\rm EW}}^p
    -\3half\beta_{4,{\rm EW}}^p  \Big]\\
    + A_{\phi\pi} \sum_{p=u,c} \lambda_p\Big[ \delta_{pu}\,(-\alpha_2 - \beta_1)
    + \alpha_4^p -\3half\alpha_{3,{\rm EW}}^p -\half\alpha_{4,{\rm EW}}^p
    + \beta_3^p -\half\beta_{3,{\rm EW}}^p -\3half\beta_{4,{\rm EW}}^p
    \Big],\label{xx2}
\end{multline}
where $A_{\phi\pi}$ and $A_{\phi\pi}$ are defined  by:
\begin{eqnarray}
A_{\phi\pi}&=&-i \sqrt{2} G_F m_\phi F_+^{B\to
\pi}f_\phi(\epsilon^*_\phi \cdot p_B);\nonumber\\
A_{\phi\pi}&=&-i \sqrt{2} G_F m_\phi A_0^{B\to
\phi}f_\pi(\epsilon^*_\phi \cdot p_B),
\end{eqnarray}
with the $B\to \phi$ form factor $A_0^{B\to \phi}=0.28$ for the
$\bar nn$ component. The Wilson coefficients $\alpha_i$ come from
vertex corrections and hard spectator corrections, and $\beta_i$
represent of contribution of annihilation diagrams, which can be
found in Ref. \cite{Beneke:2003zv}.

\begin{figure}[hbtp]
\begin{center}
\includegraphics[width=0.4\textwidth]{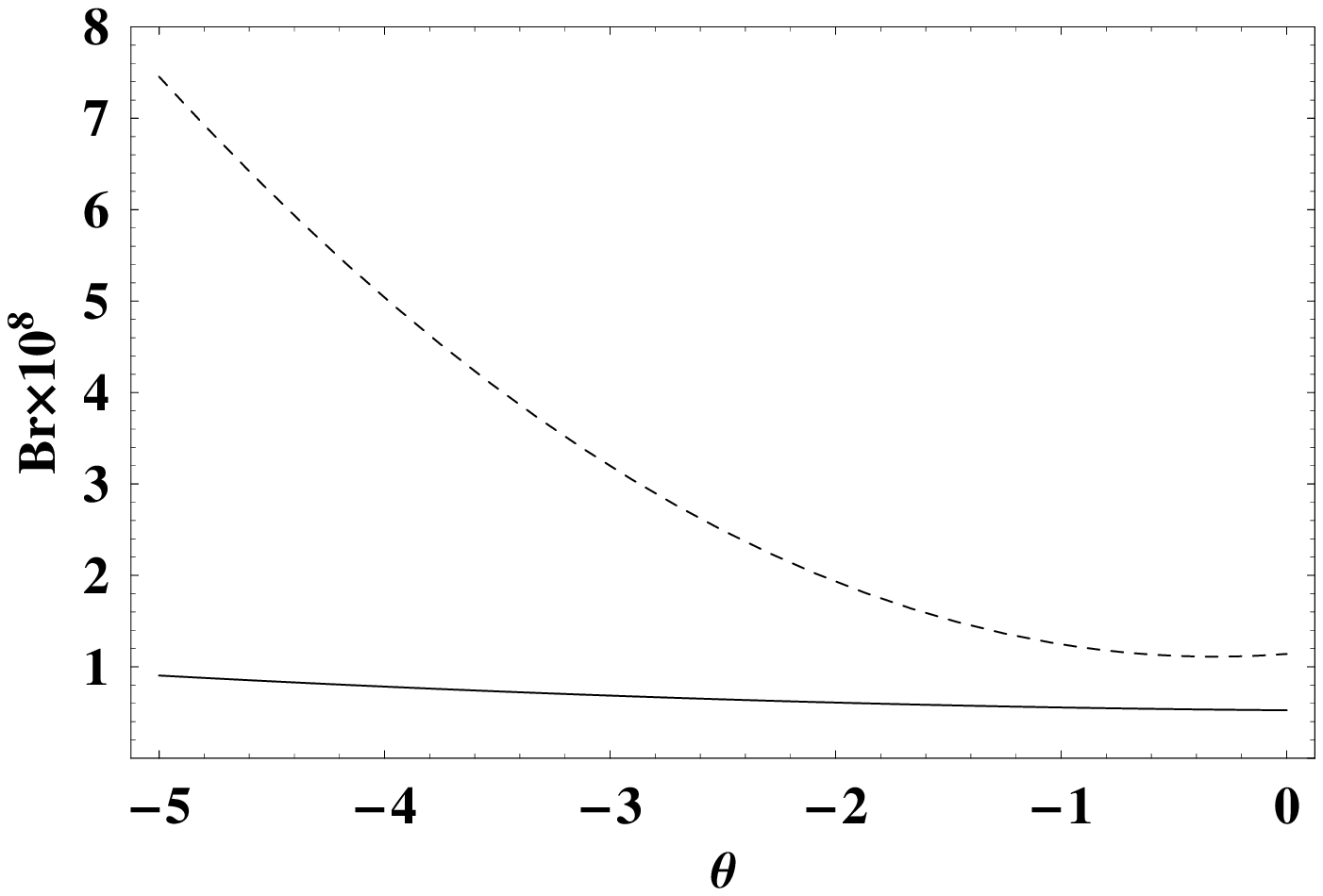}
\includegraphics[width=0.4\textwidth]{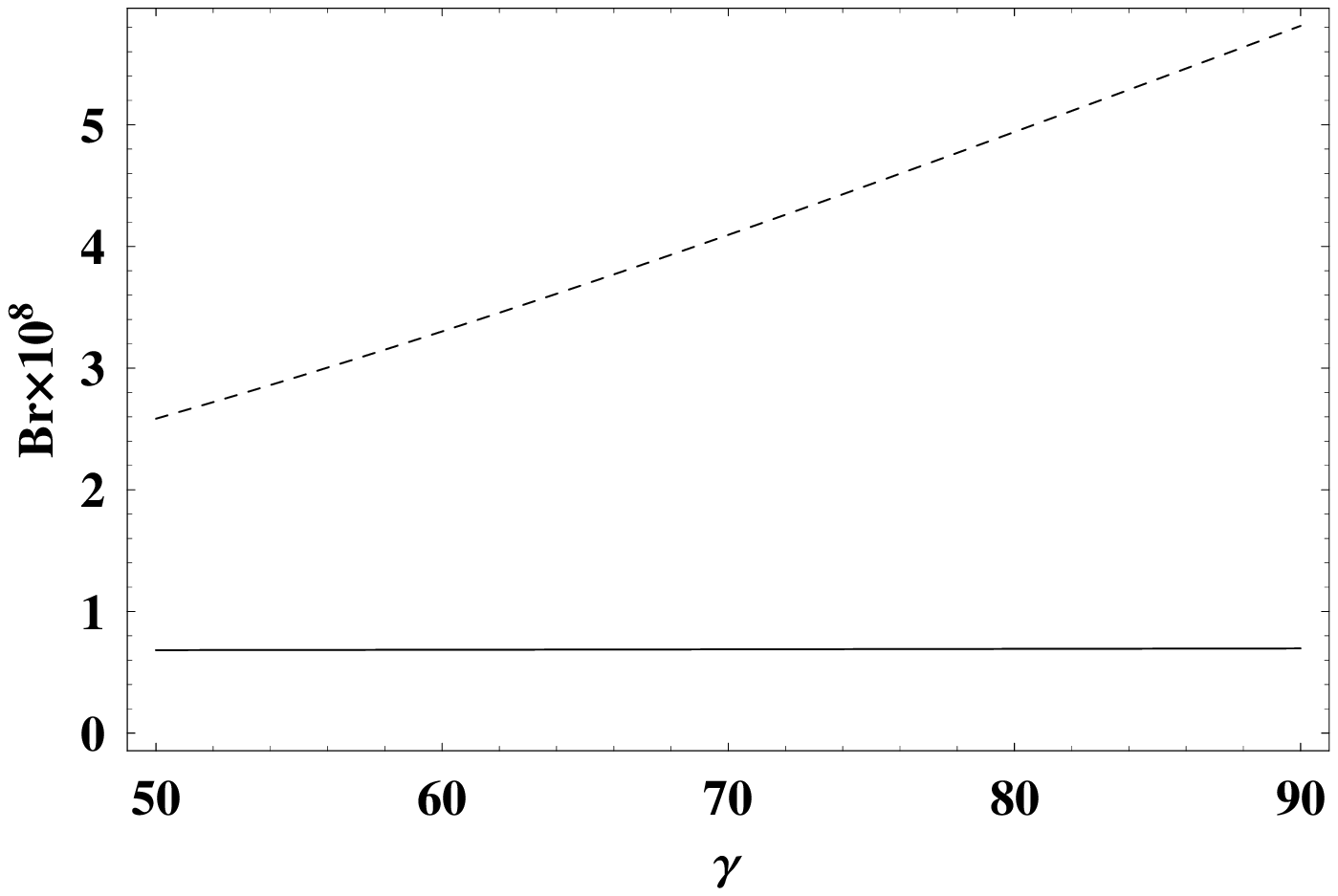}
\caption{Dependence of the CP averaged branching ratios  on the
mixing angle $\theta$ (left panel) and the CKM phase angle $\gamma$
(right panel), where the dot-dashed and  solid lines correspond to
charged channel  and neutral channel, respectively. }
\label{diagram1}
\end{center}
\end{figure}

The total amplitudes are given as:
\begin{eqnarray}
 A_{B^-\to\pi^-\phi}&=& \left[A_{B^-\to\pi^-\phi}^{QCDF} + iAbs({B^-\to\pi^-\phi})\right]\cos\theta
 + {  A}_{B^-\to\pi^-\phi}^{n\bar n}\sin\theta,\\
 A_{\bar B^0\to\pi^0\phi}&=& \left[A_{\bar B^0\to\pi^0\phi}^{QCDF} + iAbs({\bar B^0\to\pi^0\phi})\right]\cos\theta
 + {  A}_{\bar B^0\to\pi^0\phi}^{n\bar n}\sin\theta.
\end{eqnarray}
%
If one adopts the mixing angle $\theta=-3^\circ$, the branching
ratios of $B \to \phi \pi $ are:
\begin{eqnarray}\label{mixingratio}
{\rm Br}(B^- \to \phi \pi^-) &=& 3.2 \times 10^{-8}, \nonumber \\
{\rm Br}(B^0 \to \phi \pi^0) &=& 6.8 \times 10^{-9}.
\end{eqnarray}
Comparing with results in Eq.(\ref{directratio}), we found that the
branching ratio of charged channel $B^- \to \phi \pi^-$ is enhanced
remarkably. However, the $\omega$-$\phi$ mixing does not change $B^0
\to \phi \pi^0$ so much.  We plot the relations between branching
ratios and the mixing angle $\theta$, and the CKM angle $\gamma$ in
Fig.\ref{diagram1}. In the left diagram of Fig.\ref{diagram1}, we
set $\gamma=58.6^\circ$ and let $\theta$ change from $-5^\circ$ to
zero; in the right part, set $\theta=-3^\circ$ and
$\gamma\in(50^\circ, 90^\circ)$. As indicated in this diagram, the
branching ratio of $B^- \to \phi \pi^-$ is sensitive to both
$\theta$ and $\gamma$, whereas the $B^0 \to \phi \pi^0$ does not
have  this character. Physically, $B^0\to \pi^0 \phi(n\bar n)$ is a
color-suppressed process associated with angle $\gamma$, whose decay
amplitude is much smaller than color favored mode $B^\pm\to \pi^\pm
\phi(n\bar n)$. Hence, for the charged channel, the mixing
contribution becomes larger with the mixing angle $\theta$
decreasing. For the $B^0\to \pi^0 \phi$, its branching ratio is not
sensitive to these two angles because of the small amplitude of
$B^0\to \pi^0 \phi(n\bar n)$.

\begin{figure}[hbtp]
\begin{center}
\includegraphics[width=0.4\textwidth]{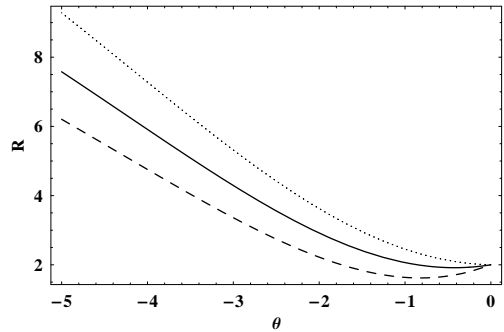}
\caption{Dependence of the ratio $R$ on the mixing angle $\theta$
with $\gamma=(58.6\pm 10)^\circ$. The solid line is the central
value of $\gamma$, while the short-dashed line and the long dashed
line correspond to the upper limit and the lower limit,
respectively. } \label{diagram3}
\end{center}
\end{figure}


Although our results are still below the experimental bound, the
branching ratio of $B^-\to \phi\pi^-$ is dramatically enhanced: to
roughly $0.6\times 10^{-7}$ if the mixing angle is $-4.64^\circ$.
This value is smaller than the upper bound only by a factor of 4. If
the future experiment reports a branching ratio of order $10^{-7}$,
it may not be the signal for any new physics at all. Since both of
the new physics scenarios and the mixing effect can give the
branching ratio of order $10^{-7}$, it is necessary to find a way to
discriminate them. We propose  a ratio $R$ of branching fractions
which is defined as:
\begin{eqnarray}\label{r}
 {{R}}=\frac{{\rm Br}(B^- \to \phi
 \pi^-)}{{\rm Br}(B^0 \to \phi
 \pi^0)}\frac{\tau_{B^0}}{\tau_{B^-}}=\left|\frac{A_{B^- \to \phi
 \pi^-}}{{A_{B^0 \to \phi
 \pi^0}}}\right|^2.
\end{eqnarray}
Without the $\omega-\phi$ mixing, $R=2$, while if the $\omega-\phi$
mixing is present, the ratio is a function of the mixing angle
$\theta$. Here we plot the relation of $R$ and $\theta$ in
Fig.~\ref{diagram3} as $\gamma=(58.6\pm10)^\circ$, through which we
can determine the mixing angle $\theta$ with the observable $R$.
From this diagram, we can obtain $R=4.3$ when $\theta=-3^\circ$ and
$\gamma=58.6^\circ$. As stated above, the neutral channel is not
changed so much, whereas the charged decay $B^\pm \to \pi^\pm \phi$
is enhanced by the mixing contribution, so the ratio $R$ is
controlled by the channel $B^\pm \to \pi^\pm \phi$. If the mixing
angle was tiny and a large branching ratio of the order $10^{-7}$ is
observed in the future, the large branching ratio must receive
dominant contributions from the new physics scenario, either enhance
the Wilson coefficients of the operators in the SM or introduce new
effective operators beyond. They will contribute to both
$B^-\to\phi\pi^-$ and $\bar B^0\to \phi\pi^0$ decays. In this case,
the ratio R is identically 2 which is dramatically different with
the one caused by the $\omega$-$\phi$ mixing effect.

\begin{figure}[hbtp]
\begin{center}
\includegraphics[width=0.4\textwidth]{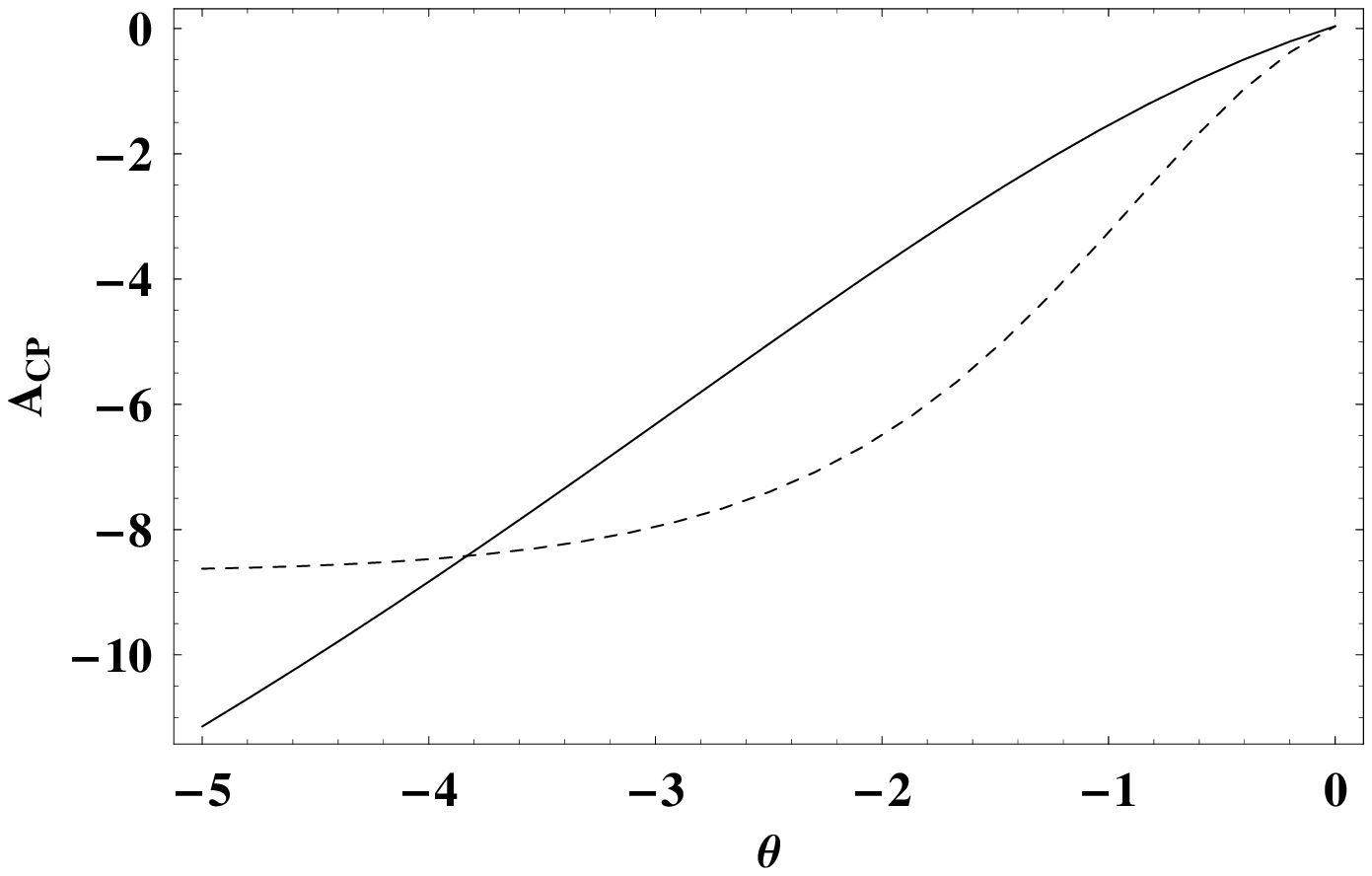}
\includegraphics[width=0.4\textwidth]{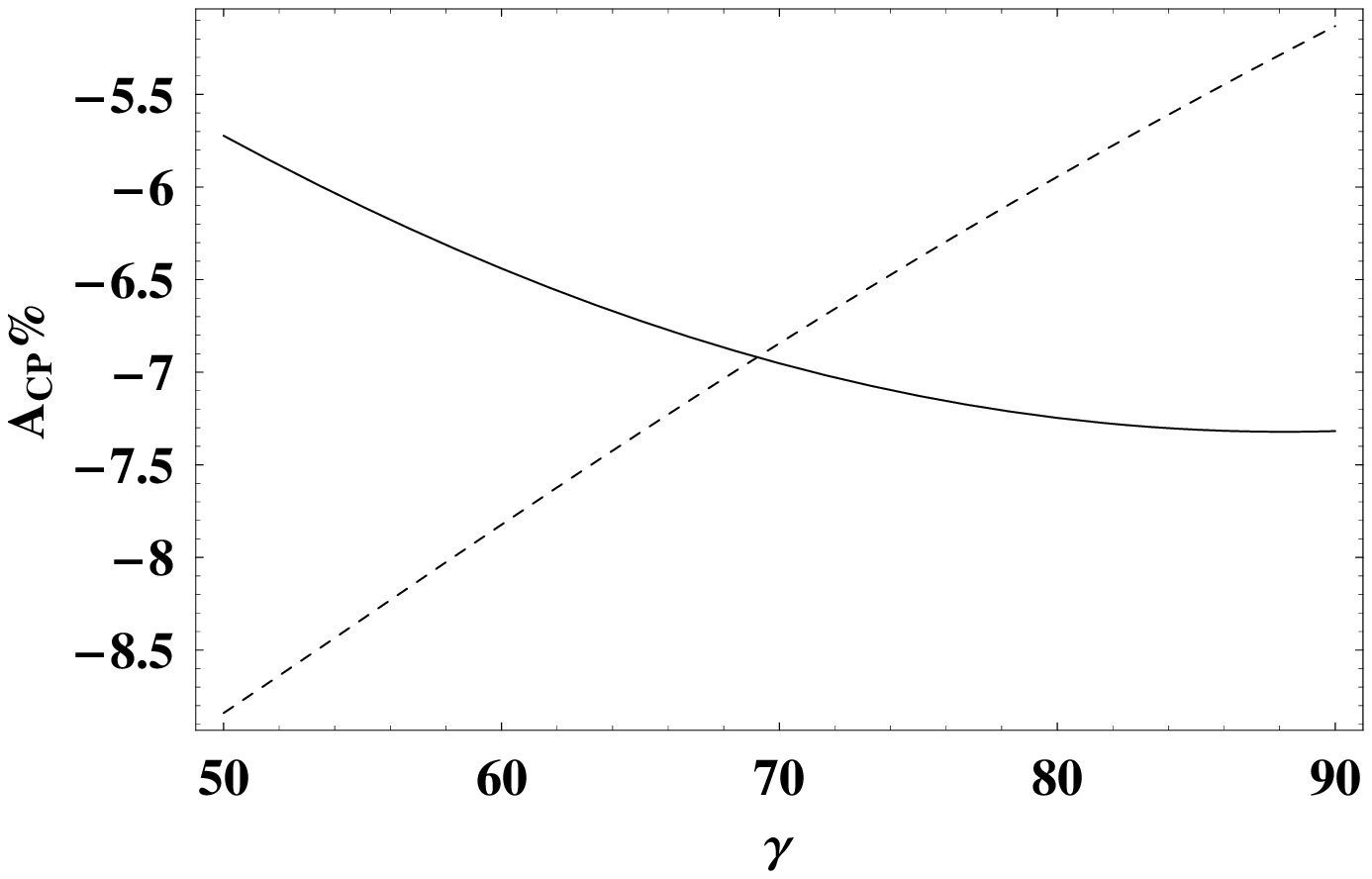}
\caption{Dependence of the direct CP asymmetries (in units of $\%$)
on the mixing angle $\theta$ (left panel) and the CKM phase angle
$\gamma$ (right panel), where dot-dashed lines and the solid lines
correspond to charged channel and neutral channel respectively.  }
\label{diagram2}
\end{center}
\end{figure}

Another observable in $B$ decays is the direct $CP$ violation
parameter, which is defined as:
\begin{eqnarray}
 A_{CP}= \frac{\Gamma(B^- \to \phi \pi^-)-\Gamma(B^
 +
\to \phi \pi^+)} {\Gamma(B^- \to \phi \pi^-)+\Gamma(B^+ \to \phi
\pi^+)}\;.\label{acp}
\end{eqnarray}
In order to have non-zero direct $CP$ asymmetry, the decay amplitude
needs to contain at least two interfering contributions with
different strong and weak phases. Since only penguin operators does
contribute to this decay mode in the absence of $\omega$-$\phi$
mixing,  the direct $CP$ asymmetry turns out to be identically zero.
In the mixing scenario, there is a small portion of the $u\bar u$
component in $\phi$ meson, and tree operators contribute so that the
direct CP asymmetries are non-zero. If $\theta<-3^\circ$, the
contribution from $n\bar n$ part dominate the $\phi\pi^-$
progressively, and the direct $CP$ violation becomes stable as
$\theta$ grows down. Because $B^0 \to \pi^0\phi(n\bar n )$ has a
small amplitude, the direct $CP$ of this decay mode comes from
interference between tree contribution of $n\bar n$ and penguin from
both $n\bar n$ and $s\bar s$, which leads to the $CP$ violation are
sensitive to mixing angle $\theta$.  With the definition in
Eq.(\ref{acp}) and the mixing angle $\theta=-3^\circ$, the direct
$CP$ violation parameters are given as:
\begin{eqnarray}\label{cpviolation}
  A_{CP}(B^- \to \phi \pi^-) &=& -8.0\%, \nonumber \\
  A_{CP}(\bar B^0 \to \phi \pi^0) &=& -6.3\%  .
\end{eqnarray}
In the left part of Fig.\ref{diagram2}, we illustrate the dependence
of $A_{CP}$ on the mixing angle $\theta$. In the right part of the
Fig.\ref{diagram2}, we set $\theta=-3^\circ$, and draw the relation
between $A_{CP}$ and the CKM angle $\gamma$.

Many uncertainties in two body non-leptonic $B$ decays, such as the
decay constants, the amplitude distributions, are constrained by
many well measured decay channels as $B \to \pi\pi,
K\pi$~\cite{data}. In the decay mode $B\to \phi\pi$, the major
uncertainties are from the mixing angle $\theta$ and the CKM phase
$\gamma$.  Then we set the CKM angle $\gamma=(58.6\pm 10)^\circ$ and
the mixing angle $\theta= -(3.0\pm 1.0)^\circ$, and get the results
with errors,
\begin{eqnarray}
{\rm Br}(B^\pm\to \phi \pi^\pm)&=& (3.2 ^{+0.8-1.2}_{-0.7+1.8})
\times 10^{-8},\nonumber\\
{\rm Br}(B^0 \to \phi \pi^0) &=& (6.8 ^{+0.3-0.7}_{-0.3+1.0}) \times
10^{-9}; \\
A_{CP}(B^\pm \to \phi \pi^\pm) &=& (-8.0^{+0.9+1.5}_{-1.0-0.1})\%, \nonumber \\
  A_{CP}(B^0 \to \phi \pi^0) &=& (-6.3^{-0.5+2.5}_{+0.7-2.5})\% ;\\
  R&=&4.3^{+1.0-1.4}_{-0.9+1.6}.
\end{eqnarray}

Our result can be directly generalized to other similar channels
such as $B\to \phi\rho$ decays. There are several differences
between $B\to \phi\rho$ and $B\to \phi\pi$ decays.  The
contributions from the mixing mechanism are larger, since the
branching ratio of $B^-\to \omega\rho^-$ is larger than that of
$B^-\to \omega\pi^-$ (in unit of $10^{-6}$): ${\rm Br}(B^-\to \omega
\rho^-)=(10.6^{+2.6}_{-2.3})
>{\rm Br}(B^-\to \omega \pi^-)=(6.9\pm0.5)$~\cite{data}. The
transverse polarization of $B\to \phi\rho$ also receives sizable
contributions from the dipole operator
$O_{7\gamma}$~\cite{Lu:2006nza}.

Because of tiny branching ratios in the SM, the authors in
Refs.~\cite{phinewphysics} argued that the decay mode $B\to \phi\pi$
is a good place for probing the new physics effect. In the present
paper, we have studied several contributions to $B\to \pi\phi$
decays in the SM. We find that the small branching fraction,
expected in the naive factorization approach, can be remarkably
enhanced by the radiative corrections and the $\omega$-$\phi$ mixing
mechanism. The final results for the branching ratio  of $B^-\to
\pi^-\phi$ is smaller than the present upper limit by a factor of
$4-20$. We conclude that, the observation of this channel with the
branching ratio of the order $10^{-7}$ may not be a clear signal for
the new physics scenario. On the contrary, that may be induced by
the $\omega$-$\phi$ mixing effect. In order to discriminate the two
different contributions, we propose to measure the ratio $R$ of the
branching fractions in the future.  The contributions from the
$\omega$-$\phi$ also provide nontrivial strong phases, which result
in large direct $CP$ asymmetries. These results can be tested on the
future experiments.

\section*{Acknowledgements}
This work is partly supported by the National Science Foundation of
China under Grant Nos. 10747156, 10625525, 10735080 and 10805037.

\end{document}